# Relationship of Quantum Entanglement to Density Functional Theory

A.K. Rajagopal and R. W. Rendell
Naval research Laboratory, Washington DC, 20375

## ABSTRACT

The maximum von Neumann entropy principle subject to given constraints of mean values of some physical observables determines the density matrix. Similarly the stationary action principle in the case of time-dependent (dissipative) situations under similar constraints yields the density matrix. The free energy and measures of entanglement are expressed in terms of such a density matrix and thus define respective functionals of the mean values. In the light of several model calculations, it is found that the density matrix contains information about both quantum entanglement and phase transitions even though there may not be any direct relationship implied by one on the other.

In an interesting paper, Wu et al [1] linked entanglement in interacting many-body quantum systems to density functional theory. They used the Hohenberg-Kohn (HK) theorem [2] on the ground state to show that the ground state expectation value of any observable can be interchangeably viewed as a unique function of either the control parameter or the associated operator representing the observable. They exhibit a relation of this to the study of quantum phase transitions. The generalization of this theory to the thermal mixed states was first given Mermin [2] which may similarly be useful in connecting the entanglement at finite temperatures near a phase transition. Mermin's work is elaborated further in [3, 4]. Here we present a generalization of these concepts for arbitrary mixed states by using the maximum entropy principle. This is equivalent to a general version of the associated minimum "free energy" principle. In this way, we establish the duality in the sense of Legendre transform between the set of mean values of the observables based on the density matrix and the corresponding set of conjugate field parameters associated with the observables. This at once implies that such quantities as Concurrence [5] or Negativity [6] that are employed to quantify quantum entanglement, which are defined in terms of the density matrix, are now shown to depend on these parameters, thus generalizing the Lemma of Wu et al [1].

When the density matrix represents a pure state, these results go over to the density functional theory given in [1].

The system density matrix $\hat{r}$ is a Hermitian, positive semi-definite, operator with unit trace, $tr\hat{r} = 1$. It is employed to determine the mean values of the $l$-th physical observable $\hat{A}_l$ defined by $\{a_l\} = tr\{\hat{r}\hat{A}_l\}$. If one of these observables is the system Hamiltonian, the corresponding constraint is just the traditional mean energy of the system, as in [1]. The information entropy associated with the density matrix is the von Neumann entropy defined as $S[\hat{r}] = -tr\hat{r}\ln\hat{r}$. The maximum entropy principle involves maximizing the von Neumann entropy with respect to the density matrix subject to the constraints of given mean values defined above and the condition of unit trace of the density matrix. This is equivalently stated in terms of the principle of minimum "free energy" defined by

$$F[\hat{r}] = tr\hat{r}\left[\sum_l \mathbf{l}_l \hat{A}_l + \mathbf{l}_0 + \ln\hat{r}\right] \tag{1}$$

The Lagrange multipliers $\mathbf{l}_l$ are the conjugate field parameters which play a dual role with the mean values $a_l$, mentioned above. The Lagrange multiplier associated with the Hamiltonian is the traditional inverse temperature in units where the Boltzmann constant is taken to be unity. The density matrix that minimizes the free energy is

$$\hat{r}_0 = \exp-\sum_l \mathbf{l}_l \hat{A}_l \Big/ Z(\{\mathbf{l}_l\}), \ Z(\{\mathbf{l}_l\}) = tr\exp-\sum_l \mathbf{l}_l \hat{A}_l, \tag{2}$$

The minimum free energy is found to be

$$F[\hat{r}_0] = -\ln Z(\{\mathbf{l}_l\}) \tag{3}$$

We then find

$$F[\hat{r}] - F[\hat{r}_0] = tr\hat{r}\{\ln\hat{r} - \ln\hat{r}_0\} \geq 0 \tag{4}$$

The expression for the free energy difference is the Kullback-Leibler relative entropy and the inequality is a version of the Jensen inequality, for $x \rangle 0$, $\ln x \geq 1 - x^{-1}$.

The HK theorem based on the density matrix is stated in terms of the maximum entropy principle or equivalently, the minimum "free energy" principle following from eq.(1). From eqs.(2, 3), we have the result

$$\frac{\partial F[\hat{r}_0]}{\partial l_l} = tr\hat{r}_0\hat{A}_l = a_l \qquad (5)$$

All these go over to the results given in [1] when the density matrix represents a pure state as happens in the zero temperature limit of the thermal density matrix.

The Lemma in [1] is thus restated in terms of the density matrix as in the traditional quantum information theory. The entanglement measure M is a functional of the density matrix, as for example, Concurrence [5] or Negativity [6]. Hence these measures all become respective functionals of the mean values.

It may be pointed out that the quantum entanglement and decoherence properties may also be time-dependent when dissipative processes are considered [7]. A time-dependent generalization of DFT exists based on a stationary quantum action principle [8] in place of maximum entropy principle discussed above for the density matrix evolution. The Legendre structure is maintained in this case as well. Thus, all these developments involve the density matrix itself and the above considerations apply here as well as in further generalizations of the Lemma in [1].

While such relationships hold, it is not clear if the functional dependencies in the free energy functional and the quantifiers of entanglement display similar characteristics. Specifically, the mathematical structure of the density matrix needed to quantify the quantum entanglement does not in general bear a linear relationship to the density matrix or to the free energy functional. Thus long-range order that determines the actual phase transition is not simply related to the entanglement characteristic residing in the density matrix. O'Connor and Wootters [9] had made the observation that a Heisenberg spin model did not exhibit a relationship between the long range order exhibited in the phase transition implied by the model and this is not evident from examining the concurrence between nearest-neighbor spin correlations. In fact, Yang [10] found in an exactly solvable quantum spin model that there is no one-to-one correspondence between quantum phase transitions and the non-analyticity property of the concurrence. In fact he surmises that some other measure of entanglement may be needed to link with the phase transition. This question has recently been investigated in some models by Cavalcanti et al [11]. Further they find that different entanglement quantifiers can indicate different orders of phase transitions. This is further supported from quantum spin model results which show that quantum phase transitions are characterized in terms of the pairwise-to-global entanglement ratio [12].

By recalling the underlying issues of phase transitions and quantum entanglement, we may infer the following: A phase transition is indicated by the onset of long-range order in the system and depends on the interactions among the constituents of the system. It is reflected as a singularity in the free energy of the system. On the other hand, defining entanglement first requires identification of discernable subsystems [13] of the overall system whose density matrix cannot be expressed as a convex sum of products of subsystem density matrices [14]. Entanglement is thus an inherent quantum "non-local" property of the system not essentially dependent on interactions among the system constituents. There may be singularities in the measures of entanglement. The examples cited above indicate that the two underlying properties are not necessarily related in the sense that the singularity in one does not imply the same singularity in the other. This is also evident from the fact that the free energy involves the system density matrix as a whole and the measures of entanglement such as Negativity involve eigenvalues of the partial transpose of the system density matrix. The two properties of the density matrix seem to be unrelated in general. We may conclude that the density matrix obtained by means of the density functional schemes considered here and elsewhere contain information about both quantum entanglement and phase transitions even though there may not be any relationship implied between the two properties.

We thank the Office of Naval Research for partial support of this work.